\documentclass[vecphys]{svmult}     
\usepackage{graphicx}        
\usepackage{epsfig}          
\begin{document}

\title*{Optical Lattices: Theory}

\author{
A.~Smerzi\inst{1}
\and
A.~Trombettoni\inst{2}
}

\institute{
Istituto Nazionale per la Fisica della Materia BEC-CRS and
Dipartimento di Fisica, Universita' di Trento, I-38050 Povo, Italy.
\texttt{smerzi@science.unitn.it}
\and
International School for Advanced Studies and Sezione INFN, Via Beirut 2/4,
I-34104, Trieste, Italy.
\texttt{andreatr@sissa.it}
}
\maketitle

\section{Introduction}

This chapter presents an overview of the properties of a
Bose-Einstein condensate (BEC) trapped in a periodic potential.
This system has attracted a wide interest in the last years,
and a few excellent reviews of the field have already appeared in
the literature (see, for instance, \cite{bloch05,jaksch05,morsch06} 
and references
therein). For this reason, and because of the huge amount of
published results, we do not pretend here to be comprehensive, but
we will be content to provide a flavor of the richness of this
subject, together with some useful references. On the other hand,
there are good reasons for our effort. Probably, the
most significant is that BEC in periodic potentials is a truly
interdisciplinary problem, with obvious connections with electrons
in crystal lattices, polarons and photons in optical fibers.
Moreover, the BEC experimentalists have reached such a high level
of accuracy to create in the lab, so to speak, paradigmatic
Hamiltonians, which were 
first introduced as idealized theoretical models to study, among others, 
dynamical instabilities or quantum phase transitions.

The key feature of our problem
is that the periodic potential naturally introduce a 
spatial {\em discreteness} in a nonlinear medium.
The periodic potential is
generally realized with two counterpropagating laser beams
\cite{anderson98,bongs00,greiner01,phillips01,
morsch01,cataliotti01,eiermann04,stoferle04,hadzibabic04},
so as to create an optical lattice (OL).
As expected in the mean-field GPE limit, the BEC Bogoliubov excitation spectrum has
a band structure,
in analogy with the electronic Bloch bands
\cite{berg98,javanainen99,choi99,chiofalo00,wu01,wu02,machholm03,menotti03,kramer03}.
When the the power of the laser is fairly larger than the chemical potential,
the lowest band dynamics maps on a
discrete nonlinear Schr\"odinger (DNLS) equation
\cite{trombettoni01}. This was an interesting remark especially
because the DNLS  was already widely investigated
{\em per se} by the nonlinear physics community
\cite{hennig99,kevrekidis01,ablowitz04} which, indeed, was immediately attracted
by the new possibilities offered by this system.

The BEC GPE dynamics in the array
can be therefore studied in the framework of
the nonlinear lattice theory
\cite{trombettoni01,abdullaev01,konotop02,alfimov02}.
The typical confining potential is given by the superposition
of an harmonic trap and a periodic potential. For a 1D OL,
the frequency at the bottom of the wells is typically of order of $\sim kHz$ in
the OL direction, and the transverse confinement is provided by the
magnetic potential (characterized by frequencies of order of $100Hz$).
The axial dynamics of a Bose condensate induced by 
an external potential with cylindrical symmetry in the transverse directions 
can be studied introducing an effective $1D$ GPE equation \cite{salasnich02}: 
for BEC in OL, assuming that the Wannier wavefunctions 
(localized in each well) can be expressed in Thomas-Fermi approximation, 
it has been shown \cite{smerzi03} that the main effect of the transverse
confinement is to modify the degree of nonlinearity of the DNLS equation,
giving raise to a generalized version of the DNLS equation.
BEC in a periodic potential can
allow for the observation of intrinsic localized modes
(i.e. matter excitations localized on few lattice sites),
as well as the study of solitons and breathers,
possibly also with condensates having a repulsive interatomic interaction. 
We should also mention that
the realization of two- and three-dimensional optical lattices
\cite{greiner01,greiner02} opens
the possibility to study discrete/nonlinear effects in higher spatial dimensions. 
A discussion of the derivation of the generalized DNLS equation is presented in Section II.

In free space, the superflow of a uniform BEC is described by
plane waves, which becomes energetically unstable 
in presence of defects when the BEC velocity is faster than sound, 
which is the Landau criterion for superfluidity.
The propagation of sound in a harmonically trapped condensate without OL
has been observed experimentally
\cite{andrews97} and studied theoretically
\cite{zaremba97,kavoulakis97,stringari98,damski04}. In the presence
of a periodic potential, the condensate wavefunction can be expanded in Bloch waves, having amplitudes
modulated with the periodicity of the OL, which can also become 
energetically unstable when the group
velocity is larger than the sound velocity 
\cite{wu01,menotti03,wu03,taylor03,boers04,kramer05-s}.
The energetic instability manifests itself with the emission of quasi-particles 
out of a condensate flowing against a small obstacle. 
This happens when the condensate velocity is larger than a critical value, which, 
in the limit of small obstacles 
is the sound velocity.  
The interplay between discreteness and nonlinearity is also
crucial for the occurrence of modulational
instabilities (MI), well known in the theory of nonlinear media.
MI are dynamical instabilities characterized by
an exponential growth of arbitrarily small fluctuations of a
carrier wave, as a result of the interplay between dispersion
and nonlinearity. The consequences
of the modulational instability of the motion of BEC wavepackets in OL have been discussed
in \cite{smerzi02}. A different parametric instability, which will not be discussed here,
can arise when modulating in time
the height of the interwell barriers or the strength of the interparticle interaction
\cite{rapti04,kramer05}.
In Section III we will discuss the excitation spectra of a BEC
in a periodic potential, while in Section IV we
review the occurrence of a discrete modulational instability,
comparing its effects with those of the Landau instability; a brief discussion of the
the propagation of sound in the OL will be also presented.
A discussion of the dynamics of BEC wavepackets in OL is given in Section V.

At last (but not at least) the high laser power
available nowadays allows for the investigation of low tunneling
rates between adjacent wells of the periodic potential:
in these regimes the quantum fluctuations play an important role,
and, with a strength of the optical potential $V_0$ large enough,
it is expected a quantum transition from
a superfluid phase to a Mott insulator phase. The model used to describe
the quantum (beyond Gross-Pitaevskii) properties of ultracold atoms in deep
optical lattices is the Bose-Hubbard Hamiltonian, which is nothing less that
the quantized version of the DNLS Hamiltonian. The basic energy scales in the Bose-Hubbard model
are the tunneling energy $K$ (which decreases by increasing $V_0$) and the 
charging energy $U_2$,
due to the interaction among
particles in the same well. Since the role of the quantum 
fluctuations depends on the
ratio $U_2/K$, the OL provide an unique way to tune the effective interaction by varying the
laser power $V_0$. The phase
structure is determined by the two competing terms of the Hamiltonian \cite{fisher89}: the
interaction energy $U_2$ leads to localization of particles in the lattice (Mott phase),
while the hopping term $K$ favors superfluidity.
The phase coherence of different condensates in the superfluid phase
(and its disappearance in the Mott regime) in the array
plays a crucial role in the dynamics, and can be experimentally 
studied observing the interference patterns created
by the condensates after turning off the trapping potential.
The observation of squeezed number states was reported in \cite{orzel01}, 
while an experimental detection of the Mott-superfluid transition has been 
reported for
3D OL in \cite{greiner02}.  A systematic study of quantum phase transitions in
low-dimensional (1D and 2D) OL is presented in \cite{kohl05}.
We also mention that adding a disordered potential - created e.g. by an optical
speckle potential \cite{lye05,clement05,schulte05} or by superimposing
laser with different periodicity \cite{roth03,fallani06} - one expects, in presence of deep optical lattices
and for large values of $U_2/K$, a Bose glass phase \cite{fisher89}. 
For space reasons, we will not discuss here the 
main properties of the Bose-Hubbard Hamiltonian and we refer the reader 
to the chapter devoted to beyond Gross-Pitaevskii effects. 

\section{Discrete Equations for the Dynamics}

The $T=0$ dynamics of a BEC in an external potential
$V(\vec{r})$ follows the GPE \cite{dalfovo99,leggett01,pethick02,pitaevskii03}
\begin{equation}
\label{GPE}
i \hbar \psi_t= - \frac{\hbar^2}{2 m}
\nabla^2 \psi + [V + g \mid \psi \mid^2] \psi,
\end{equation}
where $g=4 \pi \hbar^2 a / m$, with  $a$ the $s$-wave scattering length
and $m$ the atomic mass. The condensate wave function is
normalized to the total number of particles $N$. We write the
external potential $V=V_{\rm MT}+V_{\rm OL}$ as the sum the
optical lattice potential $V_{\rm OL}$, created by two or more 
counterpropagating laser beams, and the trap potential $V_{\rm
MT}$, whose form depends on the particular realization of the experiment. 
For a 1D OL, created by only two counterpropagating laser beams, 
it is $V_{\rm OL}(\vec{r})=V_L(y,z) \cos^2{\left( 2 \pi
x / \lambda\right)}$, where 
$\lambda=\lambda_{laser} \sin{\left( \theta/2 \right) }$, 
$\lambda_{laser}$ being the wavelength of the lasers and $\theta$ the angle
between the counterpropagating laser beams. The spacing in the
lattice is $d=\lambda / 2$ and $V_L(y,z)$ is determined by the transverse
intensity profile of the (nearly gaussian) laser beams.
E.g., in \cite{anderson98}, $\lambda=850 \, nm$ and
the $1/e^2$ radius of the transverse profile is $\approx 80 \mu m$,
an order of magnitude larger than the transverse radius of the condensate,
so that we can approximate the periodic potential by
\begin{equation}
\label{OL-potential} V_{\rm OL}(x)=V_0 \cos^2{\left( k_x x
\right)}
\end{equation}
where $k_x = 2 \pi / \lambda$,
$V_0$ is the trap depth at the center of the beam, and $V_0=s E_R$
where $E_R=\hbar^2 k_x^2 / 2m$ is the recoil energy. A 2D (3D) OL
reads $V_{\rm OL}(x,y)=V_0 \left[ \cos^2{(k_x x)} + \cos^2{(k_y
y)} \right]$ ($V_{\rm OL}(\vec{r})=V_0 \left[ \cos^2{(k_x x)} +
\cos^2{(k_y y)} + \cos^2{(k_z z)} \right]$). For the 1D periodic
potential (\ref{OL-potential}) it is useful to write
$V(\vec{r})= V_D(x)+V_L({\vec{r}})$, where $V_D(x)$ is the $x$ component of the potential $V_{\rm MT}
\equiv V_x(x)+V_y(y)+V_z(z)$. $V_D$ has a simple physical meaning:
$F = - \frac{\partial V_D}{\partial x}$ is the effective force
acting on the center of mass of a condensate wave packet moving in
the periodic potential.

When the laser power (i.e. $V_0$) is large enough,
we can use a tight-binding approximation and decompose
the condensate order parameter $\psi(\vec{r},t)$ as a sum
of wave functions $\Phi(\vec{r}-\vec{r}_j) $
localized in each well of the periodic potential:
\begin{equation}
\label{TB}
\psi (\vec{r},t)=  \sum \psi_j(t)~ \Phi(\vec{r}-\vec{r}_j),
\end{equation}
where we denote by $j$ the different wells in the array
and $\psi_j(t)=\sqrt{N_j(t)} \, e^{i \phi_j (t)}$ is the $j$-th amplitude. Normalizing to $1$
the $\Phi$'s, it follows $\sum_j \mid \psi_j \mid^2=N$. 

By replacing ansatz (\ref{TB}) in (\ref{GPE}), the
GPE reduces to a DNLS equation \cite{trombettoni01}:
\begin{equation}
i  \hbar \frac{\partial \psi_j}{\partial t} = - K
\left(\psi_{j-1}+\psi_{j+1} \right) + \epsilon_j \psi_j +
U_2\mid \psi_j \mid ^2 \psi_j,
\label{DNLS}
\end{equation}
where the tunneling rate is
\begin{equation}
K \simeq - \int d\vec{r} \, \big[ \frac{\hbar^2}{2m}
\vec{\nabla} \Phi_j \cdot \vec{\nabla} \Phi_{j+1} + \Phi_j V \Phi_{j+1}
 \big],
\label{kappa}
\end{equation}
the on-site energies are $\epsilon_n= \int d\vec{r} \, \big[ \frac{\hbar^2}{2m}
(\vec{\nabla} \Phi_n )^2+V \Phi_n^2 \big]$
and the nonlinear coefficient (which we will suppose equal in each site) is
\begin{equation}
U_2 = g N \int d\vec{r} \, \Phi_n^4.
\label{nonlinear-coeff}
\end{equation}
Naturally, if one has a 2D (3D) OL, then the ansatz (\ref{TB})
would lead to a 2D (3D)
DNLS equation.
Equation (\ref{DNLS}) is the equation of motion
$\dot{\psi_j} = \frac{\partial \cal{H}}{\partial (i \hbar \psi^\ast_j)}$,
where $\cal{H}$ is the Hamiltonian function
\begin{equation}
{\cal H} =
- K \sum \left( \psi_j \psi^\ast_{j+1} +
\psi^\ast_j \psi_{j+1} \right)
+ \sum \left( \epsilon_j \mid
\psi_j \mid^2 + {U_2 \over 2} \mid \psi_j \mid^4 \right).
\label{HAM_DNLS}
\end{equation}
Both the Hamiltonian $\cal H$ and the normalization are conserved.

In the tight-binding ansatz (\ref{TB}) one includes only
corrections from the first band, which is correct for large $V_0$.
For $V_0$ intermediate is useful to introduce contributions from
the higher bands, i.e. by considering the ansatz $\psi (\vec{r},t)=
\sum_{j,\gamma} \psi_{j,\gamma} (t)~
\Phi_\gamma(\vec{r}-\vec{r}_j)$, where $\gamma$ labels the bands:
a discussion of the resulting discrete vector equation is presented in
\cite{alfimov02}. We also notice that the DNLS equation for just
two sites describes the dynamics of BEC in a double well, which
reduces to the dynamics of a non-rigid pendulum \cite{milburn97,smerzi97,raghavan99}:
the dynamical splitting of a BEC into two parts has been
experimentally studied in \cite{shin04,schumm05}, while the direct
observation of the atomic tunneling in a BEC double well has been
reported in \cite{albiez05}.

\subsection{Effects of Transverse Confinement}

The assumption (\ref{TB}) firstly relies on the fact that
the interwell barrier $V_0$
is much higher than the chemical potentials
(e.g., in \cite{cataliotti01} for $V_0 \sim 5E_R$ it is
$\mu \sim 0.1 V_0$).  A second important condition is that
the energy of the system should be confined within the lowest band.
Higher energy bands are not contained in the 
DNLS equation, and become important
when the energy is of the order of $\hbar \omega$, where $\omega$
is the harmonic frequency of a single well of the lattice.
The effective dimensionality of the BEC's trapped in each well
can also play a crucial role \cite{pedri01,smerzi03},
by modifying the degree of nonlinearity of the DNLS equation.
In this prospect, the DNLS equation can
be seen as a zero-order (perturbative) approximation of more complicated
discrete, nonlinear equations.

The density profile of each condensate
can strongly depend on the number of atoms present at a given instant
in the same well.
This introduce site- and time- dependent
parameters in the DNLS Eq.~(\ref{DNLS}), modifying, in particular,
its effective degree of nonlinearity.
The tight-binding approximation of nonlinear systems has to be generalized 
as \cite{smerzi03}
\begin{equation}
\label{TB_gen}
\psi(\vec{r},t)= \sum \psi_j(t) ~ \Phi_j(\vec{r}; N_j(t)),
\end{equation}
with $\Phi_j(\vec{r}; N_j(t))$ depending {\it implicitly} on time
through $N_j(t) \equiv |\psi_j(t)|^2$.
We stress here, and discuss again later, that the spatial wavefunctions
$\Phi_j$ (which are considered sufficiently localized in each well)
can also depend {\it explicitly} on time
due to the excitation of internal modes.
For typical experimental setups, however, we can consider the adiabatic limit in which
the interwell number/phase dynamics is much slower that the typical
time associated
with the excitations of such internal modes (and, of course, the cases
where such modes
are not already present in the initial configuration of the system).
In this limit, which can be well
satisfied in experiments, the spatial
wavefunctions in Eq.~(\ref{TB_gen}) will adiabatically
follow the tunneling dynamics
and can be approximated with the {\it real} wavefunction
$\Phi_j(\vec{r}; N_j(t))$. A discussion of the validity of the adiabatic
approximation is in \cite{smerzi03}.

Replacing the nonlinear tight-binding approximation (\ref{TB_gen})
in the GPE (\ref{GPE}) and
integrating out the spatial degrees of freedom one finds the following discrete nonlinear
equation (DNL) \cite{smerzi03}:
\begin{eqnarray}
\label{DNLS_gen}
&& i \hbar \frac{\partial \psi_j}{\partial t} =
- \chi ~ [\psi_j(\psi^\ast_{j+1}+\psi^\ast_{j-1})+c.c.]~ \psi_j + \epsilon_j  \psi_j
+ \mu_j^{loc}~ \psi_j \cr
&& - [K+\chi~ (\mid \psi_j \mid^2 + \mid \psi_{j+1} \mid^2)] \psi_{j+1}
   - [K+\chi~ (\mid \psi_j \mid^2 + \mid \psi_{j-1} \mid^2)] \psi_{j-1}.
\end{eqnarray}
In Eq.~(\ref{DNLS_gen}),
the ``local" chemical potential is the sum of three contributions
\begin{equation}
\mu_j^{loc} = \int d\vec{r} ~ \bigg[ \frac{\hbar^2}{2m}
~(\vec{\nabla} \Phi_j)^2
+ V_L ~\Phi_{j}^2
+ g |\psi_j|^2 ~\Phi_j^4 \bigg].
\label{u}
\end{equation}
$\mu^{loc}_j$ depends on the atom number $N_j$
through the condensed wavefunction $\Phi_j$.
The tunneling rates $K_{j,j \pm 1}$
between the adjacent sites $j$ and $j \pm 1$ also depend, in principle,
on the respective populations: expanding
the wavefunctions around an average number of atoms per site, $N_0$,
and keeping only the zero
order term $\Phi_j(N_j) \simeq \tilde{\Phi}_j(N_0)$ one finds
$K_{j,j \pm 1} \approx K$, with $K$ given by Eq.~(\ref{kappa}). The relative error committed
in this approximation is order of $10^{-4}$ for typical experimental setups.
The coefficient $\chi$ is given by
\begin{equation}
\chi = - g \int d\vec{r} ~ {\tilde{\Phi}_{j}}^3 \tilde{\Phi}_{j \pm 1}.
\label{epsilon0}
\end{equation}
The on-site energies arising from any external potential
superimposed to the OL are $\epsilon_j = \int
d\vec{r}~  V_D~ {\Phi}_j^2$: $\epsilon_j \propto j^2$ ($\epsilon_j
\propto j$) when the driving field is harmonic (linear) - moreover
$\epsilon_j$ does not depend on the on-site atomic populations.
Numerical estimates show that spatial integrals involving
next-nearest-neighbor condensates, as well as terms
proportional to $\int d\vec{r}~\Phi_j^2~\Phi_{j \pm 1}^2$, can be
neglected, but not the terms proportional to $\chi$. E.g., setting 
$\zeta=g \int d\vec{r} ~ {\tilde{\Phi}_{j}}^2 \tilde{\Phi}_{j \pm 1}^2$ 
one has - 
for $V_0 \approx 20 E_R$ and $N_0 \approx 10000$ - $\chi N_0/K \sim
10^{-1}$ and $\zeta N_0/K \sim
10^{-4}$. In a double well potential e.g., with height barrier 
$V_0 \approx 2 \pi \cdot 500 Hz$ and $N_0 \approx 3000$, one has 
$\chi N_0 /K \sim 1$, while $\zeta N_0 /K \sim 10^{-3}$. For these reasons, one cannot neglect the $\chi$ terms in
Eq.~(\ref{DNLS_gen}). 
Further studies of a BEC in a double well potential 
without neglecting terms 
proportional to $\int d\vec{r}~\Phi_j^2~\Phi_{j \pm 1}^2$ 
are presented in \cite{ananikian06,kirr07}.

To make Eq.~(\ref{DNLS_gen}) useful, one has to guess the
dependence  of the localized wavefunction $\Phi_j$ on $N_j$. 
It turns out that a reasonable choice is given by supposing a
Thomas-Fermi expression for the $\Phi_j$'s: to be more explicit, 
let us introduce the potential $\tilde{V}$ at the bottom of wells, obtained 
expanding the potential $V$ around the minima. At the lowest order 
$\tilde{V} \approx (m/2) (\tilde{\omega}_x^2 x^2 +
\tilde{\omega}_y^2 y^2 + \tilde{\omega}_z^2 z^2)$. One has to compare the
interaction energy with the frequencies $\tilde{\omega}_{x,y,z}$: 
we denote by ${\cal D}=0,1,2,3$ the {\em number} 
of spatial dimensions in which one can use the
Thomas-Fermi approximation. E.g., ${\cal D}=3$ means that we can 
approximate $\Phi_j$ with the Thomas-Fermi expression 
$\Phi_j(\vec{r};N_j)\propto (\tilde{\mu}_j-\tilde{V}(\vec{r}))$ where 
$\tilde{\mu}_j$ is fixed by the normalization condition 
and depends on $N_j$ -  with ${\cal D}=2$, denoting by (let say) 
$y$ and $z$ the directions in which one can apply the Thomas-Fermi expression,
one can factorize $\Phi_j=\phi_j(x) \phi_{TF}^{(j)}(y,z)$ 
with the Thomas-Fermi expression 
$\phi_{TF}^{(j)}(y,z;N_j)\propto (\tilde{\mu}_j-(m/2) 
(\tilde{\omega}_y^2 y^2 + \tilde{\omega}_z^2 z^2)$, $\tilde{\mu}_j$ yet being 
determined by the normalization of the $\Phi_j$. Proceeding along this way, 
one gets \cite{smerzi03}
\begin{equation}
\mu^{loc}_j =  U_{\alpha} \mid \psi_j \mid^{\alpha}; \, \, \, \, \, \alpha \equiv \frac{4}{2+{\cal D}}.
\label{mu_dimensional}
\end{equation}
The coefficient $U_\alpha$ is obtained from Eq.(\ref{u}) and depends 
in general on the specific trap potential. An estimate for it 
in a particular setup is given below in Eq.(\ref{U_1}). 
The DNLS Eq.~(\ref{DNLS}) is recovered from the DNL Eq.~(\ref{DNLS_gen})
in the case ${\cal D}=0$ (i.e. $\alpha=2$) and neglecting terms proportional
to $\chi$. In conclusion the main effect of the 
transverse confinement is to change the degree of nonlinearity and
the generalized DNLS equation reads
\begin{equation}
i  \hbar \frac{\partial \psi_j}{\partial t} = - K
\left(\psi_{j-1}+\psi_{j+1} \right) + \epsilon_j \psi_j +
U_\alpha\mid \psi_j \mid ^\alpha \psi_j  - \chi {\cal F}
\label{DNLS_alpha}
\end{equation}
where ${\cal F} \equiv
[\psi_j(\psi^\ast_{j+1}+\psi^\ast_{j-1})+c.c.]~ \psi_j  +
(N_j + N_{j+1})] \psi_{j+1} +
(N_j +N _{j-1})] \psi_{j-1}$.

To make the previous result more transparent, let us consider an
harmonic trap potential $V_{\rm MT} = (m/2) (\omega_x^2 x^2 +
\omega_y^2 y^2 + \omega_z^2 z^2)$ . When the $\Phi_j$ does not
depend on $N_j$, one has ${\cal D}=0$ and $\alpha=2$, as in the
standard DNLS equation. However, for deep 1D lattices, the
effective frequencies in the $x$ direction is given by
$\tilde{\omega}_x=\sqrt{{2 V_0 k_x^2}/{m}}$ and is
$\tilde{\omega}_x \sim 10 kHz$ for $V_0 \sim 5E_R$, while
$\omega_{x,y,z}/2\pi \sim 100Hz$, for $^{87}Rb$. Then, for a
number of particles $\sim 1000-10000$ one can use a Thomas-Fermi
dependence on $N_j$ for the wavefunctions in the $y$ and $z$
directions, but not in the $x$ direction: with the previous
notation, this means ${\cal D}=2$ and $\alpha=1$. This result can
be simply obtained by factorizing the localized wavefunction
$\Phi_j$ as a product of a gaussian $\phi_j$ having width
$\sigma$ (in the $x$ direction) and a Thomas-Fermi
$\phi_{TF}^{(j)}$ (in the $y$ and $z$ coordinates): replacing in
Eq.~(\ref{GPE}) and integrating out along the $x$ direction, one obtains $\epsilon_j=\Omega j^2$,
where $\Omega=\frac{m}{2} m \omega_x^2 (\frac{\lambda}{2})^2$,
getting the DNL (\ref{DNLS_gen}) with ${\cal D}=2~(\alpha=1)$ and
\begin{equation}
U_1=\sqrt{ m \omega_r^2 g / \sqrt{2 \pi} \pi \sigma}.
\label{U_1}
\end{equation}

\section{Excitation Spectra}

In this Section  the Bloch and the Bogoliubov excitation
spectra of the system in absence of any driving field ($V_L =
0$) are derived in the tight binding
approximation. We also present a brief discussion of the comparison with numerical results
for the excitation spectra of the continuous GPE \cite{menotti03}.

\subsection{Bloch Spectrum}

The Bloch states $\Psi_p(\vec{r})=e^{ipx/\hbar} {\tilde \Psi_p}({\vec{r}})$,
where
${\tilde \Psi_p}(\vec{r})$ is periodic in the $x$ direction
with period $d$, are exact stationary
solutions of the Gross-Pitaevskii equation (\ref{GPE}). The
energy per particle $\varepsilon_{\gamma}(p)$ (Bloch energy) and the
chemical potential $\mu_{\gamma}(p)$ of such solutions form a band
structure, so that they can be labeled by the quasi-momentum $p$ and
the band index $\gamma$.

The generalized DNLS equation (\ref{DNLS_alpha}) 
describes only the lowest band of the
spectrum. Exact
solutions of the DNL equation are the "plane waves" $\psi_j= \psi_0~e^{i(k j -
\mu t)/\hbar}$, where $p=\hbar k / d$ is the quasi-momentum.
Note that the $\psi_j$ are plane waves in the lattice, but do not correspond to plane waves in real space.
Within the DNL equation framework, the energy per
particle $\varepsilon(k)$ and chemical potential $\mu(k)$
corresponding to these solutions are found to be \cite{menotti03}
\begin{equation}
\varepsilon (k)=\varepsilon^{loc} -
2~(K + 2~ \chi~N_0)
\cos \left(k\right)=\varepsilon^{loc}-
\frac{\hbar^2}{d^2 m_\varepsilon} \cos \left(k\right), 
\label{chempot1}
\end{equation}
\begin{equation}
\mu (k)=\mu^{loc} - 2~(K + 4~ \chi~N_0)
\cos \left(k\right)=\mu^{loc}-
\frac{\hbar^2}{d^2 m_\mu} \cos \left(k\right),
\label{chempot}
\end{equation}
where $\varepsilon^{loc}=2 U_\alpha
N_0^{\alpha/2}/(\alpha+2)$ and $\mu^{loc} = \mu^{loc}_j|_{\psi_l=\psi_0}= \partial (N_0
\varepsilon^{loc}) / \partial N_0$, with $N_0 = |\psi_0|^2$ the number
of atoms per well. In the previous equations we have
introduced the effective masses $m_\varepsilon$ and $m_\mu$, to
emphasize the low momenta (long wavelength) quadratic behaviour of the
Bloch energy spectrum and of the chemical potential
\cite{kramer03}. It turns out that several dynamical properties of the
system can be intuitively understood in terms of such effective
masses. This approach is quite common, for instance, in the theory of
metals, where $m_\mu \equiv m_\varepsilon$.  However in BEC, because
of the nonlinearity of the Gross-Pitaevskii equation, the two relevant
energies of the system, $\varepsilon$ and $\mu$, have the same
$\cos{(k)}$ dependence on the quasi-momentum $p$, but
different curvatures. Therefore, $m_\mu \ne m_\varepsilon$, with
\begin{equation}
{1 \over m_\varepsilon} \equiv \left.{{\partial^2
\varepsilon }\over {\partial p^2}}\right|_{0} = \frac{2 d^2 ~(K +
2~ \chi~N_0)} {\hbar^2} ,\quad
{1 \over m_\mu} \equiv \left.{{\partial^2 \mu }\over {\partial
p^2}}\right|_{0}
 = \frac{2 d^2 ~(K + 4~ \chi~N_0)}{\hbar^2}.
\label{m_e-m_mu}
\end{equation}
It is possible to extend the definition of the effective
masses to the full Brillouin zone, introducing the quasi-momentum
dependent masses
$m_\varepsilon(k) \equiv (\partial^2 \mu / \partial p^2)^{-1} =
m_\varepsilon / \cos(k)$ and
$m_\mu (k) \equiv (\partial^2 \mu / \partial p^2)^{-1} = m_\mu / \cos(k)$,
where $m_{\varepsilon} \equiv m_{\varepsilon}(0)$ and $m_\mu \equiv m_\mu(0)$. 

Similarly, one can introduce two different group velocities, defined as
\begin{equation}
\label{v_e}
{v_\varepsilon} \equiv
{{\partial \varepsilon }\over {\partial p}}
= {1 \over m_\varepsilon} {\hbar \over d}
\sin \left( k \right), \, \, \, \, \,
{v_\mu} \equiv
{{\partial \mu }\over {\partial p}}
= {1 \over m_\mu} {\hbar \over d} \sin \left( k \right).
\label{v_mu}
\end{equation}
These two different group
velocities are related by \cite{menotti03,kramer03}
$
v_\mu = v_\varepsilon + {{\partial v_\varepsilon}\over {\partial N_0}} N_0
$
with, given Eqs.~(\ref{m_e-m_mu}), $v_\mu > v_\varepsilon$. 
The current carried by
a Bloch waves with
quasi-momentum $p$ is $\rho_0~ v_\varepsilon (p)$,
where $\rho_0$ is the average particle density;
$m_\mu$, on the other hand, plays a
crucial role in the Bogoliubov spectrum, which we will discuss below.

The concept of effective mass, defined as the inverse of the
curvature of the corresponding spectrum (as that of {\it group
velocity}, defined as the first derivative) can be extended to
shallow OL, where the nonlinear tight binding approximation
breaks down.  In this case, the quasi-momentum dependence of
$\varepsilon$ and $\mu$ will not be simply described by a cosine
function, but will still remain periodic in the quasi-momentum
$p$.  In particular, the value $k$ where $m_\varepsilon(p)$
changes sign (corresponding to $\partial ^2 \varepsilon / \partial
p^2 =0$) will be greater than $\pi/2$ and will in general not
coincide with the momentum where $m_\mu$ changes sign
(corresponding to $\partial ^2 \mu / \partial p^2 =0$).

We remark that the Bloch states are not the only stationary solutions
of the Gross-Pitaevskii equation.  Because of nonlinearity, indeed,
periodic solitonic solutions can also appear for a weak enough periodic
potential, introducing new branches in the excitation
spectra \cite{machholm04}.

\subsection{Bogoliubov Spectrum}

In this subsection we study the Bogoliubov spectrum of elementary
excitations. This describes the energy of small perturbations with
quasi-momentum $q$ on top of a macroscopically populated state with
quasi-momentum $p$ [stationary solution of Eq.~(\ref{GPE})]. 

We consider
first the case $\chi=0$: in the homogeneous limit ($\epsilon_j=0$),
the stationary solutions of Eq.~(\ref{DNLS_alpha}) are plane waves
$\psi_j (t)= \psi_0  \exp{[i (k j - \nu t)]}$, with frequency
$\nu$ given by
$\hbar \nu = - 2 K \cos{(k)} + U |\psi_0|^\alpha$.
The stability analysis of such states can be carried out by
perturbing the carrier wave as
$\psi_j (t) = \left( \psi_0 + u(t) e^{i q j} + v^{\ast}(t) e^{-i q j} \right)
e^{i (k j - \nu t)}$. Retaining only terms proportional
to $u/\psi_0$ and $v/\psi_0$, one gets
\begin{equation}
i \hbar \frac{d}{dt}
\left( \matrix{
         u  \cr
         v  \cr } \right)=
\left( \matrix{
{\cal A} & {\cal C} \cr
-{\cal C}^\ast & -{\cal A} } \right)
\left( \matrix{
         u  \cr
         v  \cr } \right) = \hbar \omega_{\pm} \left( \matrix{
         u  \cr
         v  \cr } \right).
\label{gen_alpha}
\end{equation}
with ${\cal A}= 2 K \cos(k) - 2 K \cos{\left( k+q \right)} +
(1/2) U \alpha \vert \psi_0 \vert^\alpha$ and
${\cal C}=(1/2) U \alpha \psi_0^{\ast \, \alpha/2-1}
\psi_0^{\alpha/2+1}$ \cite{menotti03}.
From Eq.~(\ref{gen_alpha}) it follows that
the excitation spectrum (i.e., the Bogoliubov dispersion relation)
for the DNLS with nonlinearity degree
$\alpha$ is:
\begin{equation}
\omega_{\pm}/2K =  \sin{(k)} \sin{(q)}
\pm\sqrt{4 \cos^2{(k)} \sin^4{\left({q \over 2}\right)} +
\frac{\alpha U}{K} |\psi_0|^\alpha \cos{(k)} \sin^2{
\left({q \over 2}\right)}}.
\label{spectrum}
\end{equation}
The carrier wave becomes modulationally unstable when
the eigenfrequency $\omega$ in Eq.~(\ref{spectrum})
becomes imaginary: the condition for stability is
\begin{equation}
4K \cos^2{(k)} \sin^2{
\left({q \over 2}\right)} + \alpha U |\psi_0|^\alpha \cos{(k)} > 0.
\label{dis}
\end{equation}
When $U$ is negative (positive),
corresponding to negative (positive) scattering length, the plane waves
with $\cos{(k)}<0$ ($\cos{(k)}>0$) are stable. When the
lhs side of Eq.~(\ref{dis}) becomes negative, as a consequence
of the fact that eigenfrequency $\omega$ in Eq.~(\ref{spectrum})
becomes imaginary, there is an exponential growth
of small perturbations of the carrier wave:
we refer to this instability as the modulational instability.

In the general case $\chi \neq 0$ one can repeat the previous stability 
analysis getting \cite{menotti03}
\begin{equation}
\label{omega}
\hbar\, \omega_{\pm} \approx \frac{\hbar^2\sin(k)\sin(q)}{m_\mu d^2}
\pm 2 \sqrt{ \frac{\hbar^4\cos^2(k)\sin^4(q)}{m_\mu^2 d^4}
 +
\frac{\hbar^2 N_0}{m_\varepsilon d^2}
\frac{\partial \mu}{\partial N_0}
\cos\left(k\right)
\sin^2\left(q \right)}
\end{equation}
with the chemical potential given by $\mu = \mu^{loc} - {\hbar^2
\over d^2 m_\mu} \cos \left(k\right)$ (see
Eq.~(\ref{chempot})), and $\mu^{loc} = U_{\alpha} \mid \psi_0
\mid^{\alpha}$.  For $\alpha=2$ (i.e. ${\cal D}=0$) 
and in the limit $\chi=0$, we recover the
well known results for the discrete nonlinear Schr\"odinger equation
\cite{kivshar92}.

\section{Landau and Dynamical Instabilities}

From the relation (\ref{omega}), valid for the DNL (\ref{DNLS_alpha}) 
with $\chi \neq 0$, 
the small $q$ (large wavelength) limit of the Bogoliubov dispersion
relation becomes
\begin{eqnarray}
\hbar\, \omega \approx
{\hbar \over d m_\mu} \sin\left(k\right) \; q
+ |q| \sqrt{{1 \over  m_\varepsilon} \frac{\partial \mu}{\partial N_0} N_0
\cos\left(k\right) },
\end{eqnarray}
(we assume, for the moment, that
$\frac{1}{m_\varepsilon}\frac{\partial \mu}{\partial N_0} N_0 \cos(k) > 0$).
The linear behaviour in $q$ indicates that the system
supports (low amplitude) sound waves, propagating on top of large
amplitude traveling waves with velocity
\begin{equation}
v_{s,\pm} = \left. \hbar
\frac{\partial \omega}{\partial q}\right|_{q \to 0^{\pm}} =
\left\{\begin{array}{l l}
v_\mu + c \; ,  \hspace*{0.5cm}  (q \to 0^+) \\[1.0ex]
v_\mu - c \; ,  \hspace*{0.5cm}  (q \to 0^-) \\
\end{array}
\right.
\label{sound}
\end{equation}
where the ``chemical potential group velocity'' $v_\mu$ has been defined
in Eq.~(\ref{v_mu}), and the ``relative sound velocity'' $c$
is defined as
\begin{equation}
c = \sqrt{ {1 \over m_\varepsilon}
\frac{\partial \mu}{\partial N_0} N_0 \cos\left(k\right)}.
\end{equation}
The two velocities $v_{s,\pm}$
correspond, respectively, to a sound
wave propagating in the same and in the opposite direction of the large
amplitude traveling wave.

We remark that, contrary to the case of a Galilean invariant system
($s = 0$), the sound velocity depends on the quasi-momentum
$p$. Moreover, $v_s$ depends on the effective dimensionality of the
condensates, since (cf.~Eqs.~(\ref{mu_dimensional}) and (\ref{chempot}))
$\frac{\partial \mu}{\partial N_0} N_0 \sim \alpha~ U_\alpha~
N_0^{\alpha/2}$.  In the limit $\alpha=2$, $p\to 0$ and
$m_\varepsilon,m_\mu \to m$ we get the sound velocity in the uniform
case.

The system is energetically unstable if there exists an
$\omega<0$.  In the limit $s = 0$, this corresponds to a group
velocity larger than the sound velocity (Landau criterion for
superfluidity). When the system has a discrete translational
invariance ($s > 0$) the condition for this instability is obtained
from the Bogoliubov excitation spectrum Eq.~(\ref{omega}). Then, we
have that the system is not superfluid when $\omega < 0$,
corresponding to $v_\mu^2  >  c^2$. This result should be compared with the well known Landau 
criterion for
an homogeneous system ($s=0$), stating that the superfluid is
energetically unstable when $v^2 > c^2$, $v \equiv {{\partial
\varepsilon }\over {\partial p}} = {{\partial \mu }\over {\partial
p}}$ being the group velocity of the condensate, and $c = \sqrt{ {1
\over m} \frac{\partial \mu}{\partial N_0} N_0}$ the sound velocity.

There is a further dynamical (modulational) instability mechanism
associated with the appearance of an imaginary component in the
Bogoliubov frequencies, which disappears in the absence of interatomic
interactions, or in the translational invariant limit (if $a > 0$). The
onset of this instability in the tight binding regime, coincides with
the condition
\begin{equation}
c^2 < 0
\;\; \Rightarrow \;\;
\cos \left( {k } \right) < 0
\;\; \Rightarrow \;\;
|k| > \frac{\pi}{2}.
\label{mi}
\end{equation}
The dynamical instability drives an exponentially fast increase of the
amplitude of the - initially small - fluctuations of the condensate (while 
the energetic instability should manifest itself in polynomial time 
\cite{ianeselli06}).
Since the initial phases and amplitudes of the fluctuation modes are
essentially random, their growth induce a strong dephasing of the
condensate, and dissipates its translational kinetic energy (which is
transformed in incoherent collective and single particles
excitations).
We remark here the different scaling of the energetic and dynamical
instability with the interatomic interactions. Decreasing the
scattering length, the sound velocity decreases, and smaller and
smaller group velocities can break down the superfluidity of the system
(when $a \to 0$, the sound velocity $c \to 0$: in the limit 
of vanishing interactions the condensate is energetically unstable 
for an arbitrary small group velocity). 
On the other hand, the dynamical modulational
instability criterion does not depend on the scattering length. This
apparent paradox is simply solved noticing that the growth time of the
unstable modes actually depends on interactions, and
diverges when the scattering length vanishes ($\tau \to \infty$ when $a
\to 0$). Therefore, a noninteracting condensate is always dynamically
stable. There is a further point to remark: if we consider a
condensate moving with an increasing velocity, the system always
becomes first energetically unstable, then it hits the dynamical
instability. As a matter of fact, however, in real experiments the
energetic instability can grow quite slowly (and at zero temperature
only in presence of impurities \cite{wu01}), so that the dominant
dephasing mechanism is given by the modulational instability.

\section{Wave-Packet Dynamics}

In this Section we review the main properties to the wave-packet dynamics of a BEC
in an OL, summarizing here the results of a variational approach,
previously considered in \cite{trombettoni01,smerzi03}.
The approach uses a general variational wavefunction
\begin{equation}
\psi_j=\sqrt{\cal K(\sigma)} f\bigg(\frac{j-\xi}{\sigma}\bigg)
e^{ip(j-\xi)+i \frac{\delta}{2}(j-\xi)^2}
\label{var}
\end{equation}
where $\xi(t)$ and $\sigma(t)$ are, respectively, the center and
the width of the wavepacket, $p(t)$ and $\delta(t)$ their
associated momenta and $\cal K(\sigma)$ a normalization factor
(such that $\sum_j N_j=N$). $f$ is a generic function, even in the
variable $X=(j-\xi)/\sigma$. For simplicity, we will confine
ourself to an exponential trial wavefunction $f(X)=e^{-X^2}$ for
the standard DNLS equation (\ref{DNLS}), i.e. $\alpha=2$ and
$\chi=0$. A discussion of the general case is reported in
\cite{smerzi02}. The wave packet dynamical evolution can be
obtained by using the Euler-Lagrange equations for the Lagrangian
${\cal L}= \sum i \hbar \dot{\psi}_j \psi_j^\ast - \cal{H}$, with
${\cal H}$ given by Eq.~(\ref{HAM_DNLS}). In the following,
we rescale the time as $t \to {\hbar}t/{2 K}$,
measuring the energies in units $2K$. We also set
$\Lambda={U_2}/{2K}$
%
%
and $E_j=\epsilon_j/2K$.
The equations of motion for the variational parameters are \cite{trombettoni01}
\begin{eqnarray}
\label{p-xi}
\dot{p} &= - \frac{\partial {\cal V}}{\partial \xi};
\, \, \, \dot{\xi}=\sin{p} \, \cdot \, e^{-\eta}
\dot{\delta} = \cos{p} \Big(\frac{4}{\Gamma^2}-\delta^2 \Big)
e^{-\eta} + \frac{2 \Lambda}{\sqrt{\pi\Gamma^3}} -
8 \frac{\partial {\cal V}}{\partial \Gamma}, \nonumber
\\
&
\dot{\Gamma}= 2\Gamma \delta \cos{p} \, \cdot \,
e^{-\eta}
\end{eqnarray}
where $\Gamma \equiv \sigma^2$, $\eta = {1}/{2 \Gamma} + {\Gamma \delta^2}/{8}$
and the effective potential ${\cal V}$ is given by
${\cal V}(\Gamma,\xi)={\cal K} \int_{-\infty}^{\infty}
dn \, E_n \exp(-{2(n-\xi)^2}/{\Gamma})$. The pairs $\xi, p$ and $\frac{\Gamma}{8}, \delta$
are canonically conjugate
dynamical variables with respect to the effective Hamiltonian
\begin{equation}
\label{Hamiltonian-DNLS}
H = \frac{\Lambda}{2\sqrt{\pi \Gamma} } -  \cos{p} \, \cdot \,
e^{-\eta} + {\cal V}(\xi,\Gamma).
\end{equation}
The effective mass $m^\ast \equiv m_\varepsilon=m_\mu$ (since $\chi=0$) is given by
$\frac{1}{m^\ast} \equiv \frac{\partial^2 H}{\partial p^2}=
\cos{p} \, e^{-\eta}$: the quasi-momentum dependence of the effective mass allows a rich
variety of dynamical regimes. Solitonic solutions with a positive
nonlinear parameter $\Lambda > 0$, for instance, are allowed
by a negative effective mass.
A regime with a diverging effective mass $m^\ast \to \infty$
leads to a self-trapping of the wave packet, which has been recently experimentally
observed \cite{anker05}.

In the homogeneous lattice, only the optical potential is present
($V_{\rm MT}=V_D=0)$. Therefore the on-site energies $E_n$, as
well as ${\cal V}$, are constant. The momentum is, of course,
conserved and it is equal to the initial value:
$p(t)=p(0) \equiv p_0$.
We will discuss here only the case
$\Lambda>0$, in order to make contact with the experiments
in which $^{87}Rb$ atoms with positive scattering length $a$
are used; however, we observe that the equations of motion (\ref{p-xi}) are
invariant with respect to the replacement
$\Lambda \to - \Lambda$, $p_0 \to p_0 + \pi$ and $t \to -t$.

A detailed study of the variational equations of motion is in \cite{trombettoni01-1}.
Here we quote only the main results and we discuss rather
the physical implications and the comparison with a full numerical analysis.
This comparison is surprisingly successful in describing
even details of the quite complex dynamical and collisional behaviour.
Stability phase diagrams for such states are obtained
by inspection of the profile dynamics equations \cite{trombettoni01}.
The parameter $\Lambda$ is the ratio between the nonlinear coefficient,
induced by the interatomic interactions, and the coupling between
condensates in neighbour wells: it is the only (geometry dependent)
parameter which governs the dynamical
regimes of the system. When $\Lambda$ is small,
the wave packet spreads out; in the
opposite limit, the nonlinearity leads to a localization of the wave packet.
When $\cos{p_0}<0$, an intermediate regime arises: in this case,
the effective mass is negative and,
for a suitable values of $\Lambda$, 
a balance can be reached between nonlinearity
and diffusion. In terms of the variational parameters, this
means that in the diffusive regime, $\Gamma \to \infty$ and (if
$p_0 \neq 0$) $\xi \to \infty$, with an effective mass always finite.
On the contrary, in the self-trapped regime, $\Gamma$ remains finite
and the center of mass $\xi$ cannot go to $\infty$; furthermore,
$1/m^{\ast} \to 0$, meaning that $\eta \to \infty$ and $\delta \to
\infty$. Therefore in this regime there is an
energy transfer to the internal modes of oscillations, since $\delta$
is the momentum associated to the wave packet width: in the full
numerical solution of Eq.~(\ref{DNLS}), this corresponds to a
breakdown of the wave packet. We note that a nonlinear self-trapping
occurs also in a two-site model
\cite{smerzi97,raghavan99,ostrovskaya00}.

\begin{figure}[t]
\vspace{-0.30cm}
\begin{minipage}{5.6cm}
\caption{Plot of the wave function density $\rho_n =
\mid \psi_n \mid^2$ at times $t=0,20,40$ with $\Lambda=1$ in the
diffusive regime.
Numerical values:
$p_0=0$, $\delta_0=0$, $\Gamma_0=50$. The critical value of $\Lambda$
is in this case $\Lambda_c=24.8$. Solid lines: solutions of
Eq.~(\ref{DNLS}) with $73$ sites; dashed lines: solutions of variational
Eqs.~(\ref{p-xi}).}
\label{figure1}
\end{minipage}
\vskip-3.0cm \hfill
\includegraphics[width=3.0cm,height=6.05cm,angle=270]{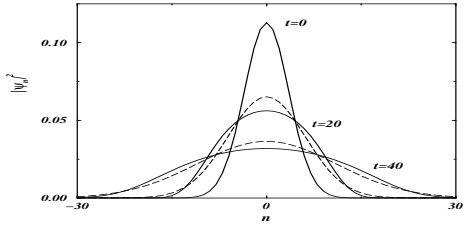}~~
\vskip-0.4cm
\end{figure}

\begin{figure}[ht]
\center
\includegraphics[width=2.75cm,height=6.25cm,angle=270]{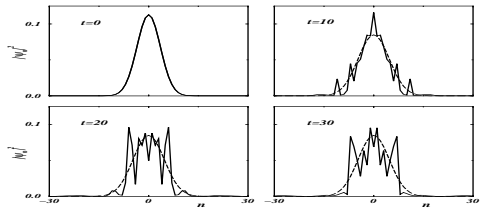}
\hskip -0.35cm
\caption{Plot of the wave function density
at times $t=0,10,20,30$ with $\Lambda=100$ in the self-trapping region.
The numerical values of the remaining parameters are as in Fig.~1.}
\label{figure2}
\vskip-0.2cm
\end{figure}

When $\cos{p_0}>0$, the solitonic regime
is forbidden and we have only the diffusive and the self-trapped regimes.
In order to show the transition between them, let us consider first
the case $p_0=0$, in which the center of mass of the wave packet does
not move ($\xi=0$).
Using as initial values $\delta_0=0$ and
$\Gamma_0 $, the initial value of the Hamiltonian
(\ref{Hamiltonian-DNLS}) is $H_0=\Lambda/2\sqrt{\pi \Gamma_0} -
e^{-1/2\Gamma_0}$. Since the Hamiltonian is a conserved quantity,
it is
$H_0=\Lambda/2\sqrt{\pi \Gamma} - e^{-1/2\Gamma-\Gamma \delta^2/8}$. 
Therefore $\frac{\Lambda}{2\sqrt{\pi\Gamma}}-H_0>0$: when $H_0>0$,
$\Gamma$ have to remain finite and the we have a self-trapped regime
in which the wave packet remains localized and the nonlinearity
forbids the diffusion. Vice versa, when $H_0<0$,
$\Gamma \to \infty$ for $t \to \infty$: the wave function spreads out
and we are in the diffusive regime.
The transition occurs at $H_0=0$, with
\begin{equation}
\Lambda_c=2 \sqrt{\pi \Gamma_0} e^{-1/2\Gamma_0}.
\label{lam-cr-p0-0}
\end{equation}
In Figures~1 and 2 we plot the density
$\mid \psi_n \mid^2$ for different
times with $\Lambda$ in the diffusive region (Fig.~1)
and in the self-trapped one (Fig.~2): the solid lines are the numerical
solutions of Eq.~(\ref{DNLS}), the dashed lines are the solutions
of the variational equations (\ref{p-xi}).
As we can see from Fig.~2, the numerical solution of
Eq.~(\ref{DNLS}) in the self-trapping region loses
its gaussian shape \cite{anker05}.
From numerical simulations is also seen that
the occurrence of the transition between the
diffusive and the self-trapped regimes does not depend on the chosen
initial conditions: what is changing is the critical value (\ref{lam-cr-p0-0}).

Also when $p_0 \neq 0$, in which the center of the wave packet
moves on the lattice, there are two distinct regimes. $H_0  > 0$,
i.e., $\Gamma(t) < \Gamma_{max}$ corresponds to the self-trapped
regime in which the boson wave packet remains localized around few
sites, while a diffusive regime occurs when $-\cos{p_0}<H_0 \leq
0$. In this case $\Gamma(t \to \infty) \to \infty$ and $\dot{\xi}
\approx -H_0/\tan{p_0} = {\rm const}$. The transition between the
regimes occurs at $\Lambda_c=2 \sqrt{\pi \Gamma_0} \, \cos{p_0} \, e^{-1/2\Gamma_0}$. With $\Lambda>\Lambda_c$,
the ratio between the initial value of the width $\sigma_0$ and the
limit width $\sigma_{{\rm max}}(t \to \infty)$ is given by
\begin{equation}
\label{alpha_max}
{\sigma_0}/{\sigma_{{\rm max}}}= ({\Lambda - \Lambda_c})/{\Lambda}.
\end{equation}
We checked the stability of the self-trapping transition also considering
different initial forms of the wave packet.
In Fig.~3 we consider a self-trapped state ($\Lambda>\Lambda_c$):
the variational prediction is that $\dot{\xi} \to 0$ and that
$\xi \to const$. As time progresses, the width increases
(and it goes asymptotically to a constant value) and the momentum conjugate
to the width goes to infinity. The full numerical solution
cannot go to this state, because the transfer of energy to the internal state
breaks down the wave packet:
when the average position approaches to value predicted from
the variational analysis (thick line), the wave packet deforms until it 
breaks. In the inset we compare the numerical and the variational average position,
where this deformation determines a deviation between
the two lines. We observe that, despite the fact that the
variational analysis cannot
exactly follow the full dynamics in the self-trapping, it can, however, 
predict the occurrence of the transition and give a fairly accurate 
estimate of the
critical point.

\begin{figure}[t]
\vspace{-0.30cm}
\begin{minipage}{6.0cm}
\caption{Density profiles at times $t=0,1.25,2.5,3.75$ (solid lines)
and at $t=5$ (dotted) for $p_0=\pi/4$, $\Gamma_0=100$ and
$\Lambda=50$ ($\Lambda_c=24.8$). The thick line represents the
asymptotic value predicted from the variational analysis.
In the inset the variational (dashed line) and numerical (solid)
average position vs. time are plotted.}
\label{figure3}
\end{minipage}
\vskip-3.0cm \hfill
\includegraphics[width=3.0cm,height=5.7cm,angle=270]{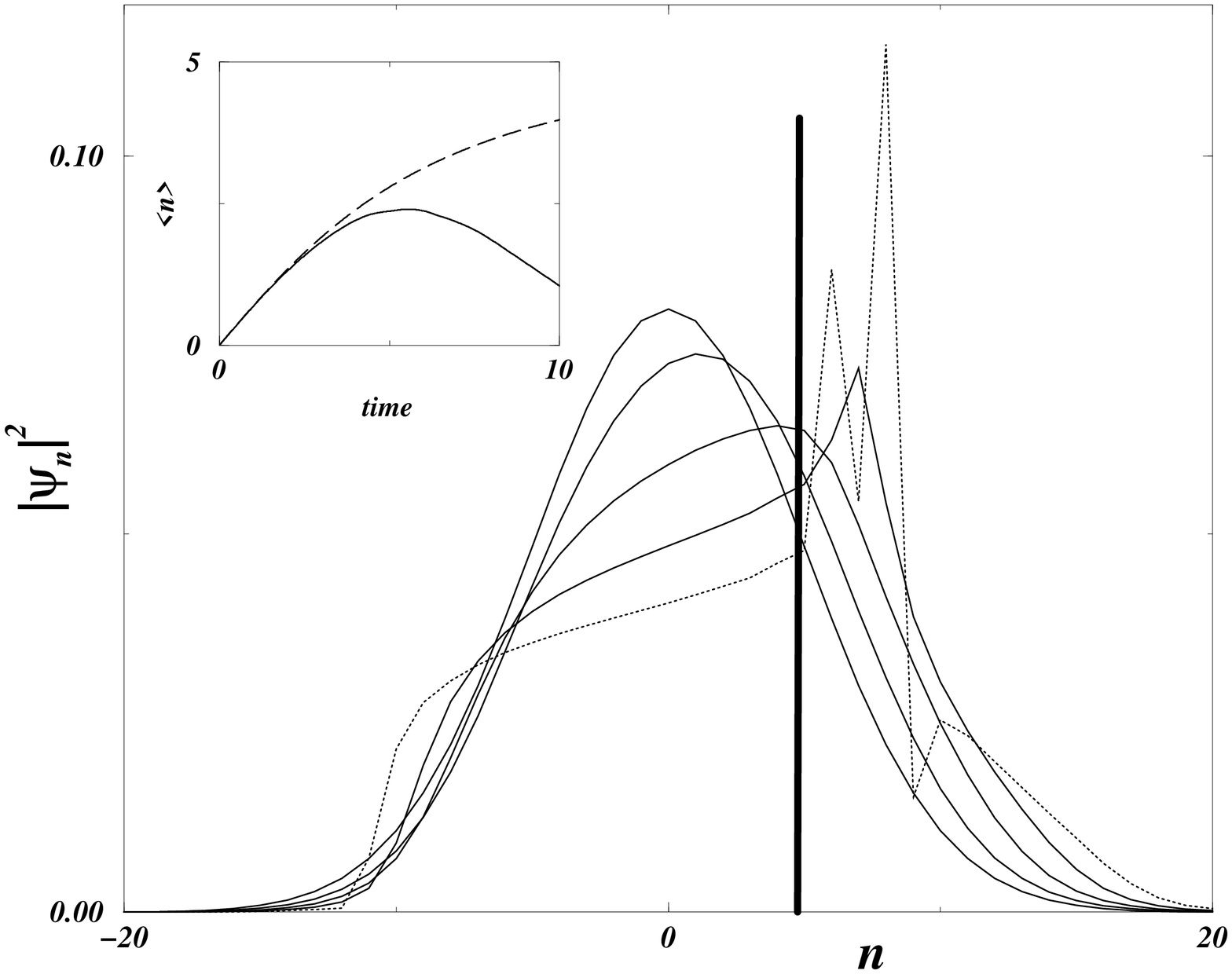}~~
\vskip-0.4cm
\end{figure}

For $\cos{p_0}<0$, soliton-like structures are present
(cf.~\cite{hennig99,kevrekidis01,ablowitz04} for more references
on discrete solitons and intrinsic localized excitations).
When $\cos{p}<0$
the self-trapping condition is given by
$H_0>\mid \cos{p_0} \mid$ and the critical value is
\begin{equation}
\Lambda_c=2 \sqrt{\pi \Gamma_0} \mid \cos{p_0} \mid \,
(1-e^{-1/2\Gamma_0}).
\label{lam-cr-cos-neg}
\end{equation}
For $\Lambda<\Lambda_c$, $\Gamma \to \infty$, while
for $\Lambda>\Lambda_c$, $\Gamma$ remains finite. A soliton solution can be determined by imposing
$\dot{\Gamma}=\dot{\delta}=0$. One finds \cite{trombettoni01}
\begin{equation}
\Lambda_{sol} = 2 \sqrt{{\pi}/{\Gamma_0}} \mid \cos{p_0} \mid
\, e^{-1/2\Gamma_0}.
\label{lam-sol}
\end{equation}
For $\Lambda=\Lambda_{sol}$ the center of the wave packet moves with
a constant velocity $\dot{\xi}$ and its width remains
essentially constant in time. We observe that for $\Gamma_0>1$, it is
$\Lambda_c<\Lambda_{sol}$.
In Fig.~4 we plot
the average position and the width for $\Lambda=\Lambda_{sol}$.
Since we are not using periodic boundary conditions,
when the wave packet arrives to the end of the lattice, it hits
a wall and upon rebounding, it regains its original shape. 
For $\Lambda_c<\Lambda<\Lambda_{sol}$, $\xi \to \infty$ while
$\Gamma(t)$ oscillates, corresponding to a breather solution.
When $\Gamma_0>1$, the breather region extends until
$\Lambda_{breath}>\Lambda_{sol}$ \cite{trombettoni01-1}.

\begin{figure}[t]
\vspace{-0.30cm}
\begin{minipage}{5cm}
\caption{Width (dotted line) and average position (solid line)
calculated numerically for $\Lambda=\Lambda_{sol}$ and $p_0=3 \pi
/4$ in a finite array of $73$ sites.}
\label{figure4}
\end{minipage}
\vskip-1.95cm \hfill
\includegraphics[width=6.0cm,height=2.25cm]{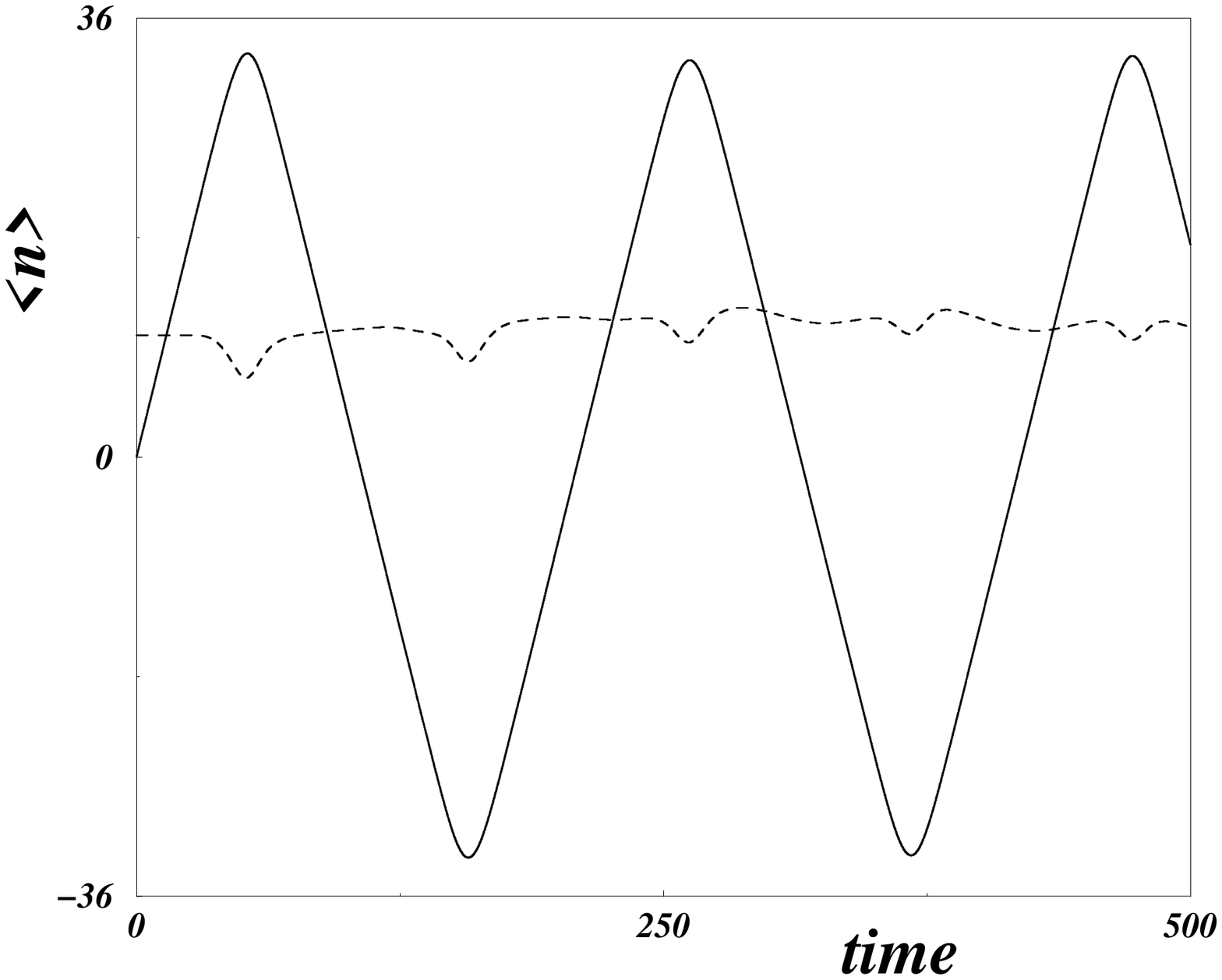}~~~~~~~~
\vskip-0.4cm
\end{figure}

Before concluding, we discuss the relation between
the present results and the modulational instability.
As we discussed, if we consider a
small perturbation on a plane wave $\psi_j \propto e^{ip_0j}$,
stability analysis shows that
when $\cos{p_0}<0$ the eigenfrequencies of the linear modes become
imaginary driving an exponential growth of small perturbations.
When $\cos{p_0}>0$, the plane wave is stable.
In the present case, we are considering not a plane wave, but a
localized wave function; therefore, we may expect to find the result
previously stated by considering the case $\Gamma >> 1$. In this case,
when $\cos{p_0}>0$, $\Lambda_c \to \infty$ and the self-trapped region
disappears: this corresponds to the result which there is no
modulational instability for $\cos{p_0}>0$. To the contrary, when
$\cos{p_0}<0$, then $\Lambda_c \to 0$ and always the system exhibits
instability to small perturbation: this means that we have only 
self-trapping, as expected.

To conclude this Section, we observe that the variational approach
can be applied also in vertical arrays (i.e., when the gravity is
acting) when Bloch oscillations occur. Similarly, Bloch
oscillations are also possible in horizontal optical lattices
realized by two counterpropagating laser beams with a frequency
detuning varying linearly in time \cite{morsch01,fallani04}. The
DNLS description of the dynamics is confined to the first band,
and then a complete description of the Bloch oscillations in a
tilted potential requires the study of the continuous GPE (see
\cite{cristiani02,wu03,jonalasinio03,breid06} and references therein).
A discussion on the
Landau-Zener tunneling is presented in the following chapter.
In an harmonic trap, one can induce and study dipole
oscillations suddenly moving the magnetic potential: if the
initial trap displacement is smaller than
a critical value, it is possible to observe coherent
Josephson-like oscillations \cite{cataliotti01}. When the initial
displacement is larger than a critical value, the modulational
instability \cite{smerzi02} breaks down the dipole oscillations
\cite{cataliotti03}.
\\
\\
{\bf Acknowledgments:} It is a pleasure to thank
our colleagues and friends with whom we had pleasant brainstormings.
Among the members of the BEC group in Trento, special thanks go to
I. Carusotto, F. Dalfovo, S. Giorgini, C. Menotti,
L.P. Pitaevskii and S. Stringari. Stimulating discussions 
with M. Albiez, T. Anker, 
J. Esteve,  R. Gati, and M. Oberthaler are gratefully acknowledged.
We also thank L.A. Collins, A.R. Bishop, P.G. Kevrekidis,
D.J. Frantzeskakis, S.R. Shenoy, G. Giusiano, 
F.P. Mancini and P. Sodano
and the members of the experimental group at LENS
(Florence): F.S. Cataliotti, C. Fort,
M. Inguscio, F. Minardi, G. Modugno and M. Modugno.


\end{document}